\documentclass[preprint,sort&compress]{elsarticle}
\usepackage{pifont}
\usepackage{natbib}
\usepackage{geometry}
\usepackage{fleqn}
\usepackage{graphicx}
\usepackage{txfonts}
\usepackage{amssymb}
\usepackage{subfig}
\usepackage{color}
\usepackage{ulem}
\usepackage{hyperref}
\usepackage{miller}
\usepackage{pdflscape}

\begin{document}

\begin{frontmatter}
\title{Statistical Study of the Defect Cluster Morphology in the Primary Damage 
of Tungsten from Collision Cascades from Five Inter-atomic potentials}

\author[barc,Hbni]{M Warrier}
\ead{manoj.warrier@gmail.com}

\author[barc]{U Bhardwaj}
\ead{haptork@gmail.com}
        
\address[barc]{Computational Analysis Division, BARC, Visakhapatnam, 
Andhra Pradesh, India -- 530012}
        
\address[Hbni]{Homi Bhabha National Institute, Anushaktinagar, Mumbai,
Maharashtra, India -- 400094}

\begin{abstract}
The size and morphology of defect clusters formed during primary damage play a 
crucial role in the subsequent microstructural evolution of irradiated 
materials. Molecular dynamics (MD) simulations of collision cascades in 
tungsten (W) were performed using five interatomic potentials (IAPs): the 
quantum-accurate machine-learned Spectral Neighbor Analysis Potential (W-SNAP), 
the machine learning-based tabGAP potential, and three embedded-atom method 
(EAM) potentials. A total of 3,500 MD simulations were conducted with primary 
knock-on atoms (PKAs) at energies of 5, 10, 20, 50, 75, 100, and 150 keV. PKAs 
were launched in 100 random directions at each energy to ensure statistical 
validity. Analysis was performed using CSaransh , a web-based tool for 
large-scale collision cascade databases, to quantify: (i) the number of defects 
(isolated and clustered), (ii) defect cluster morphologies, (iii) defect 
cluster size distributions and (iv) the number of sub-cascades formed. We show 
that the difference in the formation energy of self interstitial atom dumbells 
along the \hkl<1 1 0> and \hkl<1 1 1> directions critically influence defect 
cluster morphology. This impacts both primary damage and subsequent damage 
evolution. Our results indicate that IAP stiffness and interaction range 
independently  do not affect defect count. However, these parameters—combined 
with defect formation energies, threshold displacement energies, and other 
factors—significantly influence defect production. Notably, stiffer IAPs 
exhibit a tendency to form more sub-cascades at PKA energies below 60 keV.
\end{abstract}


\end{frontmatter}

\section{Introduction} \label{Introduction}
Molecular dynamics (MD) has been widely used to study primary radiation damage 
in materials (\cite{Stoller2012293} and references therein). Energetic ions or 
neutrons create primary knock-on atoms (PKAs) that initiate collision cascades. 
During these cascades, numerous atoms are initially displaced from lattice 
positions. This cascade of displaced atoms peaks within a pico-sec and 
recombines with local vacancies to create the "primary damage" state 
(containing interstitials, vacancies, and defect clusters). This primary damage 
evolves over time through continued irradiation, defect migration, mutual 
defect interactions, and reactions with pre-existing microstructures 
\cite{KaiHistRev2019,DomainOKMC2004,MFRT-KMC-2008}. \\

Several researchers have investigated interatomic potential (IAP) effects on 
collision cascades and residual defects. Byggm\"astar et al. demonstrated in 
iron that adjusting the stiffening range of pair potentials—accounting for 
short-range atomic interactions—can reproduce experimental threshold 
displacement energies ($E_D$) \cite{EffectZBLEDisp}. Their work further 
revealed that high-dose damage evolution depends on defect cluster 
formation/stability, governed by the equilibrium component of interatomic 
potentials. Sand et al. applied two levels of pair-potential stiffening to 
Marinica et al.'s tungsten embedded atom method (EAM) potential \cite{MS}. They 
have compared the results for collision cascades from these potentials with 
three other IAP and documented variations in: Self-interstitial atom (SIA) 
counts relative to Frenkel-pair formation energies, Vacancy migration energies, 
Displacement energies, and  Stiffness-to-range ratios as defined in 
\cite{PhysRevB.66.134104}, while also examining stiffening effects on defect 
distributions. Becquart et al. have analyzed collision cascades using seven 
empirical tungsten IAPs and five iron IAPs 
\cite{BECQUART2021152816,DEBACKER2021152887}. \cite{BECQUART2021152816} shows 
clear correlations between the threshold displacement energy (TDE), replacement 
collision sequences (RCS) and the quasi static drag (QSD) along \hkl<1 1 0> 
direction of these bcc crystals. They recommend careful reproduction of \hkl<1 
1 0> atomic interactions during potential stiffening. In 
\cite{DEBACKER2021152887} they study the effects of potential stiffening on 
cascade properties like single interstitial and vacanciy counts, fraction of 
interstitials and vacancies in clusters, the maximum sizes of defect clusters 
and the correlation of these with cascade morphology, defined by sphericity and 
volume of the cascades. \\

Bhradwaj et al. have classified the defect clusters observed in primary damage 
based on the morphology including orientation of the defect clusters 
\cite{BhardwajSavi}. MD simulations of collision cascades with three different 
EAM based IAP revealed similar classes of defect clusters 
\cite{Bhardwaj3IAPCmp}. However, significant variations exist in the quantities  
of defect clusters across these morphological classes for the three IAPs. The 
stability, transport and interactions of the defect clusters depend critically 
on their size and morphology. For example, the stability of \hkl<1 0 0> defect 
clusters varies with both size and internal structure 
\cite{Bhardwaj100Stability}. It is reasonable to assume that this principle 
likely extends to other morphological classes. Since these factors govern 
microstructural evolution, identifying which IAP yields defect morphology 
distributions matching experimental observations is essential. Unfortunately no 
experimental techniques currently exist to identify the various morphologies of 
defect clusters and quantify them at these resolutions. Moreover, for 
comparison with experiments, in-situ experiments need to be carried out, since 
the primary damage state evolves with time and at high radiation doses cause 
overlapping cascades at the same location. \\

Over the past decade, quantum-mechanical density functional theory (DFT) 
calculations have generated extensive databases of atomic configurations, 
including total energies, atomic forces, and stresses. This has enabled machine 
learning (ML) methods to construct advanced IAPs 
\cite{Behler2016,MishinMLPot,BartokMLPot}. Wood and Thompson have developed a 
spectral neighbor analysis potential (WSNAP) for tungsten which aims to improve 
the accuracy and quality of the MD predictions in order to approach the 
accuracy of DFT \cite{Wood2017SNAPPot}. Separately, Bygm\"astar et al. have 
developed a tabulated Gaussian approximation potential (tabGAP) for 
Mo-Nb-Ta-V-W alloys which shows high accuracy when applied to multicomponent 
alloys, but is less accurate for single element simulations 
\cite{BygmastartabGAPAlloy}. This study compares collision cascade defect 
formation in tungsten simulated using the ML based IAP (WSNAP and tabGAP) and 
the three EAM based IAP, namely DND-BN \cite{DND-BN}, MS \cite{MS}, and JWPot 
\cite{JWPot}. We conducted 3,500 MD simulations with primary knock-on atoms 
(PKAs) at energies of 5, 10, 20, 50, 75, and 150 keV, launched in 100 random 
crystallographic directions per energy. This expands substantially on our prior 
work \cite{Bhardwaj3IAPCmp} which explored only these three non-ML potentials 
with limited PKA energies and directional sampling. We present results on the 
defects count (single and in clusters), the maximum defect cluster sizes, 
defect cluster morphology and IAP-dependent variations in these properties. \\

The following section describes our molecular dynamics (MD) simulation 
methodology for collision cascades and provides a brief comparative analysis of 
the five interatomic potentials. Section 3 presents results on, defect counts 
(isolated and clustered), defect cluster size distributions, morphological 
classifications of defect clusters, quantification of defects within each 
morphological class, sub-cascade formation statistics, etc. We conclude with an 
integrated discussion of these findings and their implications. 

\section{Description of the MD Simulations of collision cascades}
\label{DescMD}
The Large Scale Atomic/Molecular Massively Parallel Simulator (LAMMPS) was used 
to carry out the MD simulations \cite{LAMMPS}. Different cubic simulation boxes 
of the sizes specified in Table.\ref{Table_1}, were initially subjected to a 
NPT relaxation for 10 ps at 300 K and 0 bar pressure using periodic boundary 
conditions (pbc) along all the three directions. The time step used in the NPT 
simulations is one fs. Simulation box sizes were designed to confine cascades 
within a central region excluding the outermost three unit cells. 
Post-processing verification confirmed this containment: all cascades remained 
within the designated volume except for rare channeling incidents. \\

Following relaxation, a central atom in the simulation box wasdesignated as the 
PKA. For each PKA energy specified in table.\ref{Table_1}, we performed 100 
collision cascade simulations using an NVE ensemble. The PKA was directed 
toward 100 randomly generated points on a unit sphere centered at its position, 
with the initial velocity of the PKA corresponding to the PKA energy. 
Electronic stopping was modeled as a frictional term as described in 
\cite{HarshEStopLAMMPS}. The outermost unit cells of the simulation box are 
fixed and the adjacent two unit cells to the outermost cells are temperature 
controlled at 300 K. Variable time stepping is used along with PBC along all 
three directions during the NVE simulations which last up to 20 ps. The 
positions of atoms displaced above a threshold value and their velocity 
components are output with a specified frequency during the simulations. At the 
end of the simulation, all atom positions and their potential energy are output 
for post-processing.\\

\begin{table}[ht]
\centering
\caption{The simulation cell size and number of atoms used at each PKA energy
in the MD simulations}
\label{Table_1}
\begin{tabular}{|c|c|c|c|} \hline
S.No & PKA Energy (keV) & Simulation Size & No. of Atoms \\ \hline
  1   & 5   & 50 $\times$ 50 $\times$ 50 & 250000 \\ \hline
  2   & 10  & 75 $\times$ 75 $\times$ 75 & 843750 \\ \hline
  3   & 20  & 85 $\times$ 85 $\times$ 85 & 1228250 \\ \hline
  4   & 50  & 100 $\times$ 100 $\times$ 100 & 2000000 \\ \hline
  5   & 75  & 125 $\times$ 125 $\times$ 125 & 3906250 \\ \hline
  6   & 100 & 150 $\times$ 150 $\times$ 150 & 6750000 \\ \hline
  7   & 150 & 200 $\times$ 200 $\times$ 200 & 16000000 \\ \hline
\end{tabular}  
\end{table}

\subsection{Brief comparison of the inter-atomic potentials used}
\begin{figure}[h]
\centering{\includegraphics[width=\textwidth]{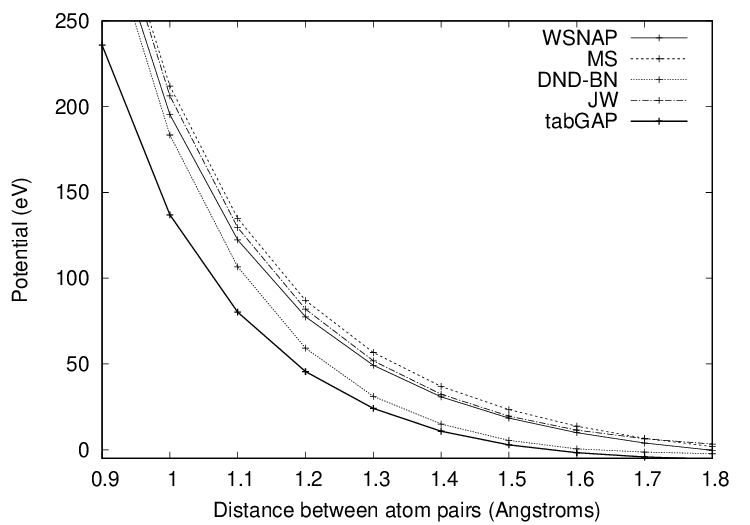}}
\caption{\label{Fig1} Total potential energy of a pair of atoms as a function 
of the distance between them.}
\end{figure}

The above described MD simulations were carried out using two ML based 
potentials, WSNAP \cite{Wood2017SNAPPot} and tabGAP 
\cite{BygmastartabGAPAlloy}, and three EAM based potentials, DND-BN 
\cite{DND-BN}, MS (specifically the $MS_h$ potential) \cite{MS} and JWPot 
\cite{JWPot}. Fig.\ref{Fig1} illustrates pair potential variation with atomic 
separation. Table.\ref{Table_2} summarizes key stiffening parameters for IAP, 
where, r1 and r2 are transition distances for smooth interpolation between 
original EAM and ZBL potential, R is the range, defined as the inter-atomic 
separation for which the pair potential is 30 eV, and S is the stiffness, 
defined as the slope of the potential at R. The two ML potentials, WSNAP and 
tabGAP, were fitted to the ZBL potential and to a screened Coulomb potential 
respectively and therefore did not require stiffening.

\begin{table}[ht]
\centering
\caption{The pair potential stiffening parameters for the five IAP. In the
second and third columns for the WSNAP and tabGAP potentials, ZBL and SC
represent the pair potentials these two ML based potentials have been fitted
to, with SC denoting a Screened Coulomb potential.}
\label{Table_2}
\begin{tabular}{|c|c|c|c|c|c|} \hline
IAP		& r1 (\AA)	& r2 (\AA)	& R (\AA) & S	& S/R ($\AA^{-1}$) \\ \hline
JW		& 1.39		& 2.6		& 1.4	& -150	& -107		\\ \hline
DND-BN	& 1.1		& 2.25		& 1.3	& -230	& -177		\\ \hline
MS		& 1.3		& 2.0		& 1.45	& -137	& -95		\\ \hline
WSNAP	& ZBL		& ZBL		& 1.4	& -145	& -104		\\ \hline
tabGAP	& SC		& SC		& 1.27	& -223	& -176		\\ \hline
\end{tabular}  
\end{table}

Table.\ref{Table-3} is a comparison of various properties of W as obtained from 
the five IAP with Experiments and DFT results. The following properties, which 
have implications on the primary damaged satate have been calculated:
\begin{itemize}
	\item The lattice parameter (a),
	\item coefficients of the stiffness tensor (C11, C12, C44),
	\item Bulk modulus (B) and Shear modulus (S), 
	\item Poisson ratio ($\nu$),{}
	\item vacancy formation energy (VFE),
	\item Formation energies of self interstitial atoms (SIA) in Octahedral, 
	tetrahedral, dumbbell (DB)$\hkl<100>, \hkl<110>, \hkl<111>$ configurations, 
	triangular defect clusters (SIA-tri), 
	\item difference in formation energies of the $\hkl<110>$ and $\hkl<111>$ 
	dumbells,
	\item minimum of the Energy vs volume curve,
	\item coefficient of thermal expansion ($\alpha$) and
	\item the threshold displacement energies (mean, maximum and minumum)
\end{itemize}

\begin{landscape}
\begin{table}[h]
\centering{}
\caption{Table of comparison of the five interatomic potentials with 
experiments and DFT results. The percentage of relative error with respect to 
either experiments or to DFT, whichever is lesser, is given in brackets. 
Relative errors in excess of 20 $\%$ is made bold-face for easy 
identification.}
\label{Table-3}
\begin{tabular}{|l|c|c|c|c|c|c|c|} \hline
Property          & DFT    & Expt.  & JW                  & MS                   & DND-BN              & WSNAP               & tabGAP        \\ \hline 
a ($\AA$)         & 3.185  & 3.165  & 3.160 (0.16)        & 3.160 (0.16)         & 3.17 (-0.16)        & 3.18 (-0.47)        & 3.18 (-0.47) \\\hline
C11	(GPa)         & 521    & 522    & 494.46 (5.09)       & 619.76 (-18.96)      & 535.60 (-2.80)      & 518.77 (0.43)       & 543.21 (-4.26)\\\hline
C12	(GPa)         & 195    & 204    & 200.44 (-2.79)      & 255.88 ({\bf-31.22}) & 205.13 (-5.19)      & 195.50 (-0.26)      & 186.52 (4.35) \\ \hline
C44 (GPa)         & 147    & 161    & 152.08 (-3.46)      & 192.40 ({\bf-30.88}) & 159.73 (-8.66)      & 144.72 (1.55)       & 137.86 (6.22) \\ \hline
B (GPa)           & 315    & 310    & 298.45 (3.73)       & 377.18 ({\bf-21.67}) & 315.29 (-1.71)      & 303.26 (2.17)       & 305.42 (1.48) \\ \hline 
S (GPa)           & 166    & 161    & 150.05 (6.80)       & 188.22 (-16.91)      & 161.93 (-0.58)      & 151.49 (5.91)       & 154.06 (4.31) \\ \hline  
$\nu$             & 0.28   & 0.28   & 0.29 (-3.57)        & 0.29 (-3.57)         & 0.28 (0.0)          & 0.27 (3.57)         & 0.26 (7.14) \\ \hline
VFE (eV)          & 3.36   & 3.8    & 3.84 (-14.29)       & 3.56 (-5.95)         & 3.65 (-8.63)        & 3.23 (3.87)         & 3.40 (-1.19) \\ \hline
SIA-Oct (eV)      & 12.27  & 10.5   & 11.93 (-13.62)      & 11.70 (-11.43)       & 10.28 (2.10)        & 11.67 (-11.14)      & 11.67 (-11.14) \\ \hline
SIA-Tet (eV)      & 11.72  & 11.2   & 11.24 (-0.36)       & 10.99 (1.87)         & 10.31 (7.95)        & 11.22 (-0.18)       & 11.94 (-6.61) \\ \hline
SIA-DB-100 (eV)   & 12.2   & 11.5   & 12.19 (-6.00)       & 11.50 (0.0)          & 10.17 (11.57)       & 10.94 (4.87)        & 11.93 (-3.74) \\ \hline
SIA-DB-110 (eV)   & 10.58  & 10.6   & 10.15 (4.06)        & 9.83 (7.09)          & 10.01 (5.39)        & 9.84 (6.99)         & 10.72 (-1.32) \\ \hline
SIA-DB-111 (eV)   & 10.29  & 10.3   & 9.91 (3.69)         & 9.54 (7.29)          & 9.57 (7.00)         & 9.76 (5.15)         & 10.56 (-2.62) \\ \hline
DB 110-111 (eV)   & 0.29   & 0.30   & 0.24 (17.24)        & 0.29 (0.0)           & 0.44 ({\bf-51.72})  & 0.08 ({\bf72.41})   & 0.15 ({\bf48.28})  \\ \hline
$\alpha (10^{-6} K^-1)$ & 4.9 & 4.5 & 8.74 ({\bf-78.32})  & 5.19 (-5.87)         & -0.97 ({\bf119.83}) & 7.14 ({\bf-45.72})  & 6.79 ({\bf -38.59}) \\ \hline
TDE mean (eV)     &        & 90.0   & 133.79 ({\bf48.65}) & 96.79 (7.54)         & 100.57 (11.75)      & 111.79	({\bf24.21}) & 109.85 ({\bf 22.06}) \\ \hline
TDE max	(eV)      &        &        & 249.0               & 199.70               & 163.00              & 220.70              & 187.70 \\ \hline
TDE min (eV)      &        &        & 54.3                & 43.00                & 42.30               & 45.70               & 38.30 \\ \hline
TDE 100 (eV)      &        & 42.0   & 54.30 ({\bf-29.29}) & 43.00 (-2.38)        & 42.30 (-0.71)       & 45.70 (-8.81)       & 38.30 (8.81)\\ \hline
\end{tabular}
\end{table}
\end{landscape}

\section{Results \label{Results}}

\begin{figure}[h]
\centering{\includegraphics[width=\textwidth]{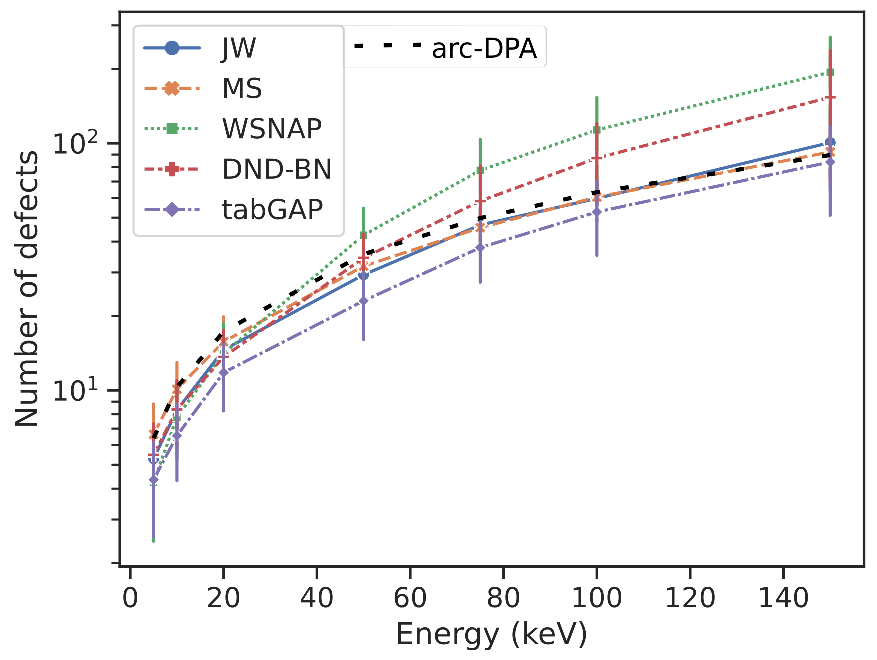}}
\caption{\label{Fig2} Plot of the number of defects from the five IAP as a 
function of PKA energy for the PKAs launched in 100 random directions. The 
standard deviation is also shown in the figure.}
\end{figure}

Fig.\ref{Fig2} shows defect count distributions versus PKA energy for all five 
potentials. All interatomic potentials (IAPs) exhibit broad defect count 
distributions across PKA energies, reflecting both crystallographic direction 
dependence and inherent statistical variations from atomic vibrations. For 
reference, tungsten arc-DPA values \cite{KaiNatureComm2018} are superimposed. 
The key observations from the figure are: defect count distribution width 
increases with PKA energy, IAPs showing elevated mean defect counts at lower 
PKA energies ($<$30 keV) do not maintain this trend at higher energies, JW and 
MS potentials demonstrate close quantitative agreement with each other and the 
tabGAP potential consistently yields the fewest defects across all energies. \\

\begin{figure}
\centering{\includegraphics[width=\textwidth]{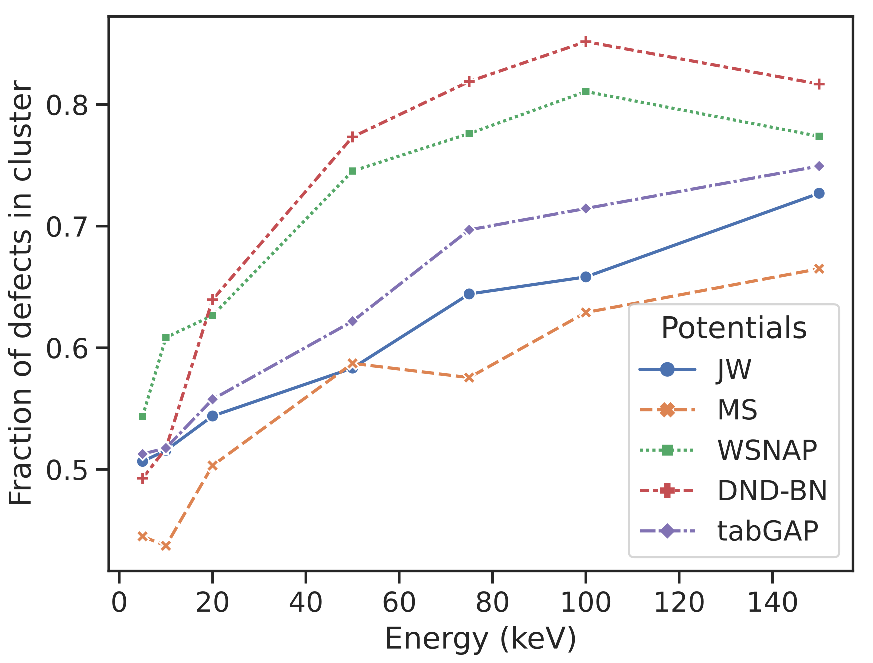}}
\caption{\label{Fig3} Fraction of defects in clusters as a function of PKA
energy for the five inter-atomic Potentials.}
\end{figure}

Fig.\ref{Fig3} shows the clustered defect fraction as a function of PKA energy 
for the five IAPs. A consistent hierarchy emerges across nearly all 
energieswith the DND-BN potential showing the maximum fraction and the MS 
potential showing the least. Note that the DND-BN potential has the lowest 
defect formation energy (summing up the values in rows 10-14 in 
Table\ref{Table-3}) which could be a factor for this. However the hierarchy 
observed does not correspond to this logic and there are other factors too that 
influence cluster formation in collision cascades. At a PKA energy of 150 keV 
the fraction of defects in clusters varies from 0.65 to 0.8, whilst at 5 keV 
PKA energy the fraction varies from 0.45 to 0.55. \\

\begin{figure}[h]
\centering{\includegraphics[width=\textwidth]{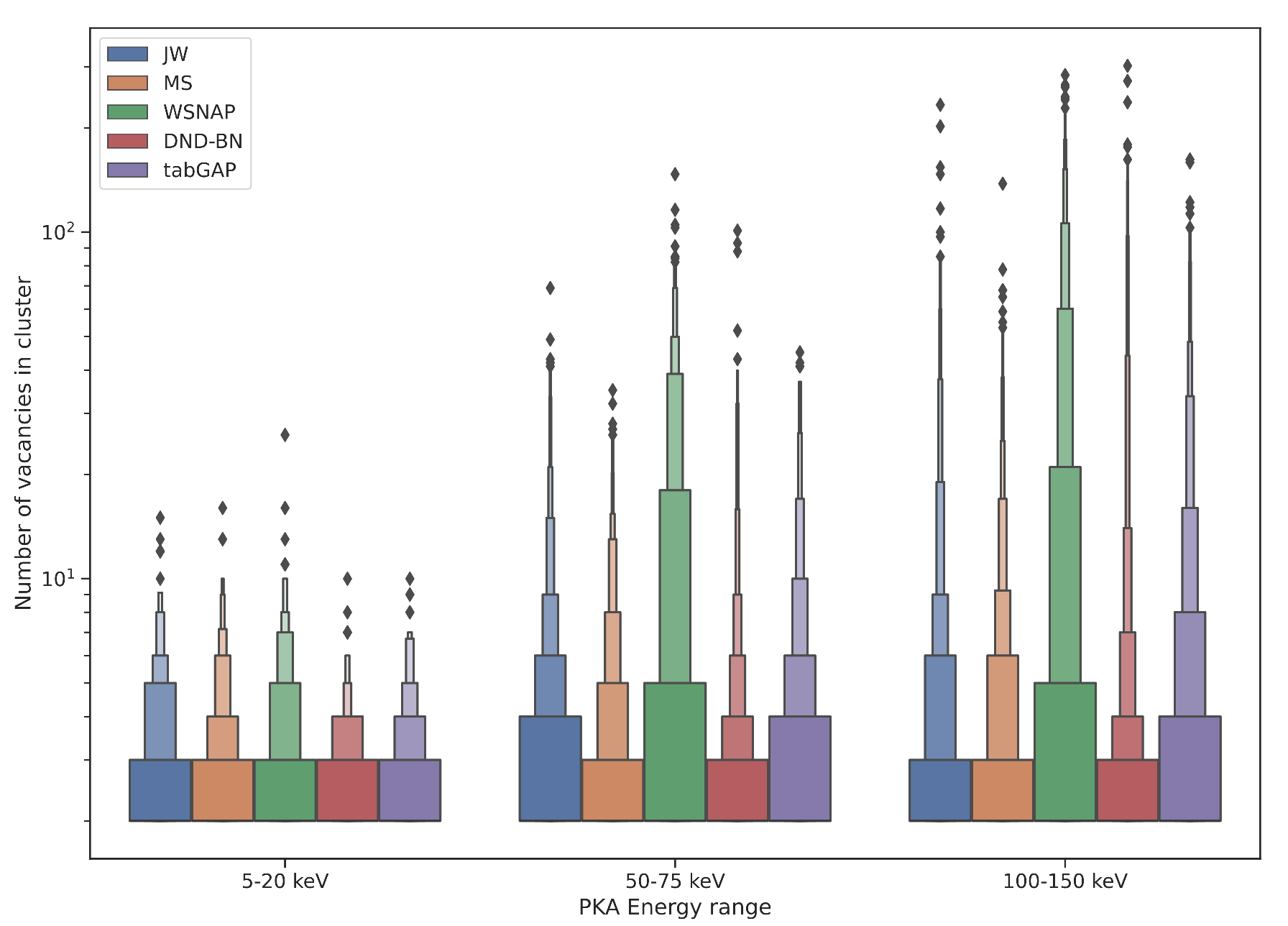}}
\caption{\label{Fig4} Box plots of the size of vacancy clusters from the 5 IAP 
for three different sets of PKA energies, namely 5-20 keV, 50-75 keV and 
100-150 keV.}
\end{figure}

\begin{figure}[h]
\centering{\includegraphics[width=\textwidth]{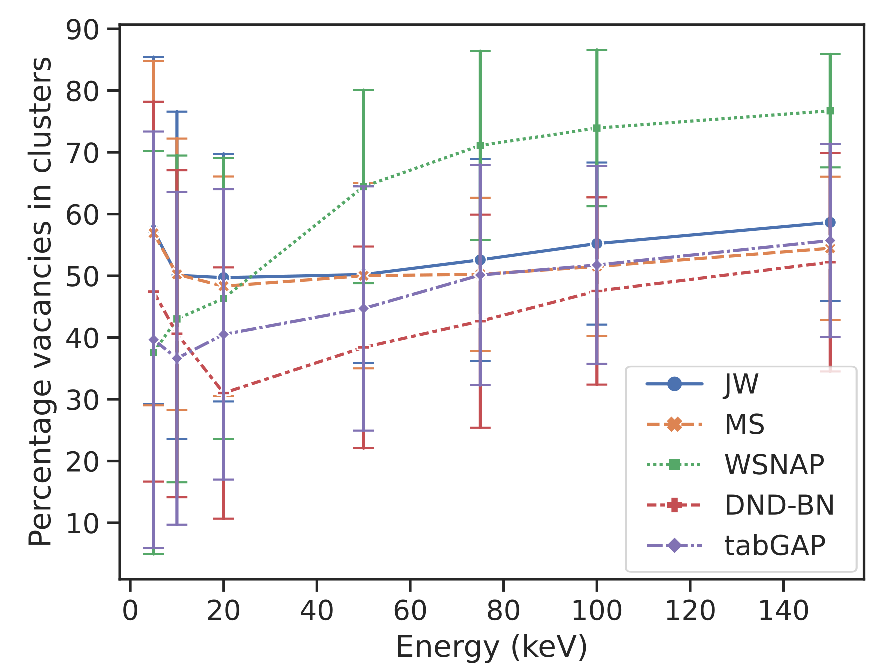}}
\caption{\label{Fig5} In cluster vacancies}
\end{figure}

The vacancy cluster size distributions from the five IAP are shown in 
Fig.\ref{Fig4} for PKA energies in the range 5-20 keV, 50-75 keV and 100-150 
keV. The percentage of vacancies in clusters are shown as a function of PKA 
energy in Fig.\ref{Fig5}. The standard deviation from the 100 random directions 
are also shown in the figure. It is seen that the WSNAP potential shows the 
largest percentage of vacancies in cluster at PKA energies greater than 20 keV, 
which aligns with it having the lowest vacancy formation energy 
(table.\ref{Table-3}). \\

\begin{figure}
\centering{\includegraphics[width=\textwidth]{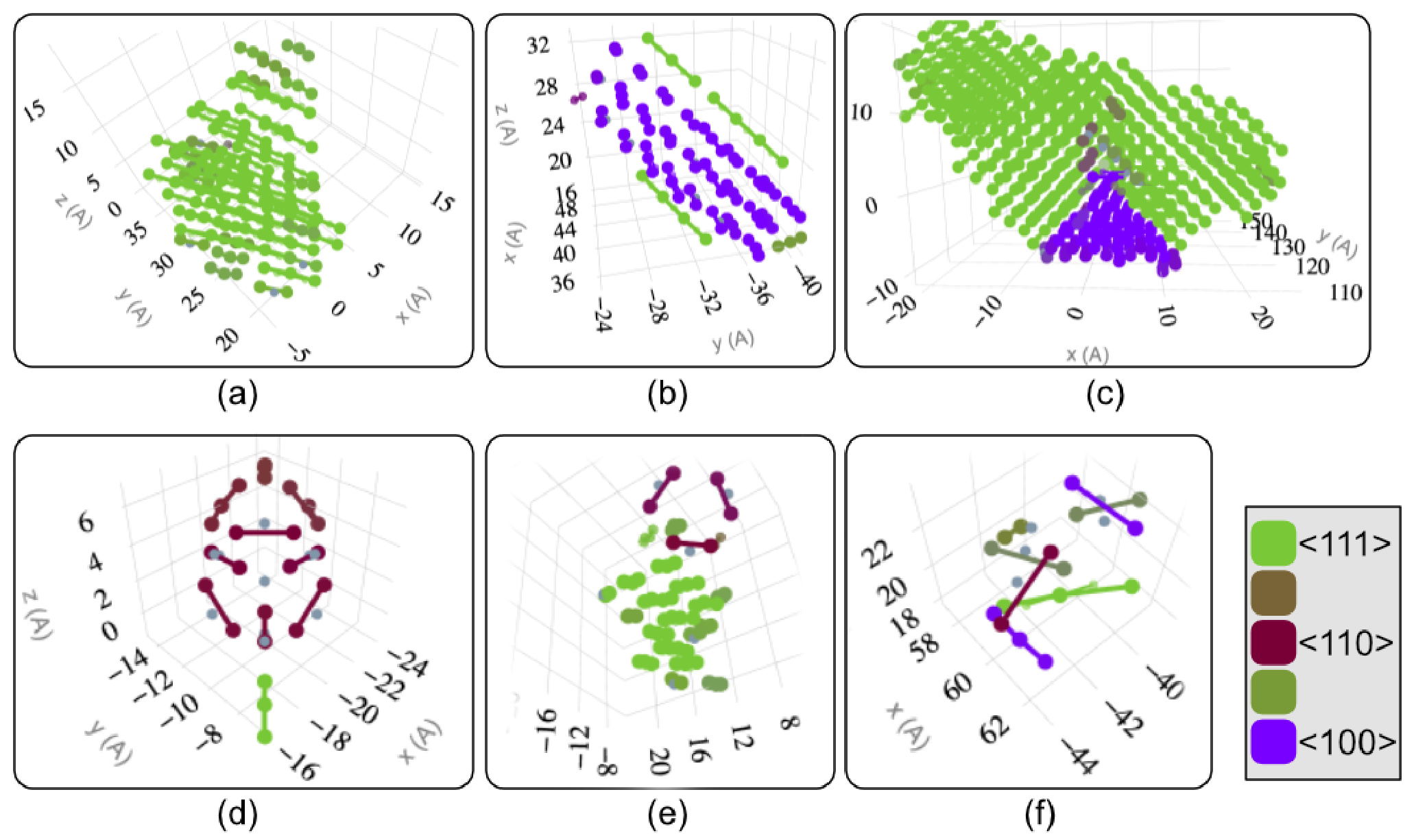}}
\caption{\label{Fig6} The six classes of defects seen in W which are observed
in the primary damage at the end of collision cascade simulations. Note that
these are present irrespective of the inter-atomic potentials used 
\cite{BhardwajSavi}.}
\end{figure}

Fig.\ref{Fig6} shows the six basic defect cluster classes seen in the primary 
damage state in all the 5 IAP using graph based analysis \cite{BhardwajSavi}.
They are listed below along with the ascii symbols used to define them:
\begin{enumerate}
\item	Fig.\ref{Fig6}.(a) represents 1/2\hkl<1 1 1> dislocations and clusters
        (ascii symbol $||$-111),
\item	Fig.\ref{Fig6}.(b) represents \hkl<1 0 0> dislocations and clusters 
        with \hkl<1 1 1> and \hkl<1 1 0> crowdions bounding them (ascii symbol
        $||$!),
\item	Fig.\ref{Fig6}.(c) represents mixed 1/2\hkl<1 1 1>--\hkl<1 1 0>--\hkl<1 
0 0> dislocations (ascii symbol $||$//),
\item	Fig.\ref{Fig6}.(d) represents C15 like rings (ascii symbol @),
\item	Fig.\ref{Fig6}.(e) represents C15 like rings with \hkl<1 1 1> clusters
        (ascii symbol $||$@), and
\item	Fig.\ref{Fig6}.(f) small random arrangement of crowdions (ascii symbol
        \#).
\end{enumerate}

\begin{figure}
\centering{\includegraphics[width=\textwidth]{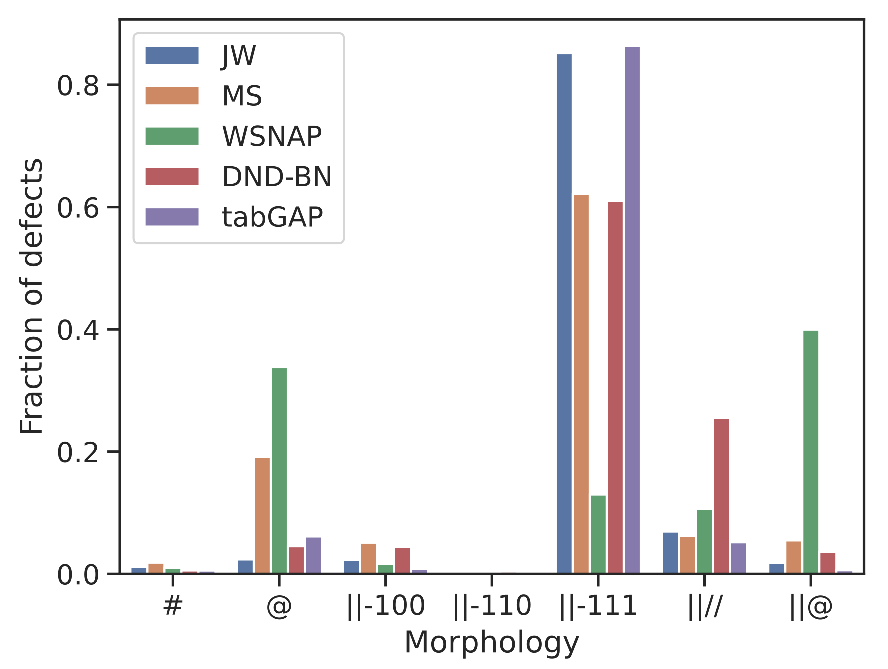}}
\caption{\label{Fig7} The Fraction of defects in each of the morphologies for
five inter-atomic Potentials.}
\end{figure}

Figure \ref{Fig7} presents the morphological distribution of defect clusters in 
the six primary defect cluster classes seen in tungsten. An extra cluster type, 
\hkl<1 1 0>, is included since a few such defect clusters were also observed 
during analysis. The 1/2\hkl<1 1 1> defects dominate (>80 \%) in JW/tabGAP 
potentials and constitute ~60 \% in MS/DND-BN. In contrast, WSNAP exhibits only 
~15\% \hkl<1 1 1> defects, with C15-like ring structures prevailing (35\% pure 
rings, 40\% ring-\hkl<1 1 1> hybrids). The MS potential also shows moderate 
ring formation (20\% C15 rings, 5\% hybrids). This trend correlates with both 
potentials’ training on liquid tungsten configurations 
\cite{MS,Wood2017SNAPPot}. Notably, Fig. \ref{Fig4} reveals hexagonal ring 
components aligned along \hkl<1 1 0> directions. While MS and WSNAP show 
similarly low \hkl<1 1 0> defect formation energies (Table \ref{Table-3}), 
WSNAP’s significantly higher \hkl<1 1 1> formation energy compared to the MS 
potential, favors C15-like clustering. Thus, the relative and absolute 
formation energies of \hkl<1 1 0> versus \hkl<1 1 1> defects critically govern 
C15-like cluster prevalence.

\begin{figure}
\centering{\includegraphics[width=\textwidth]{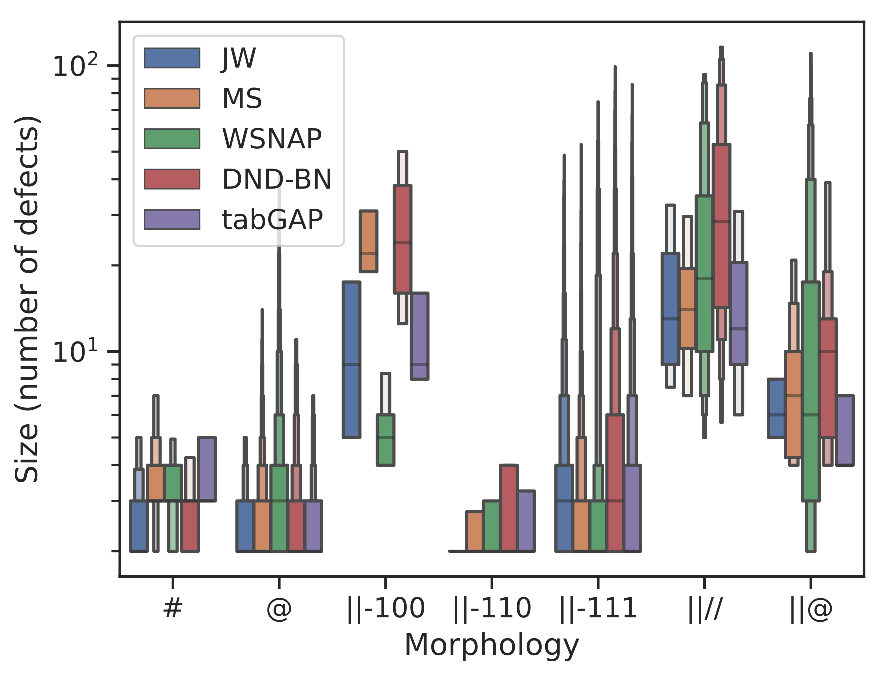}}
\caption{\label{Fig8} The defect size distribution in each of the morphologies
for five inter-atomic Potentials.}
\end{figure}

The distribution of the sizes of the defect clusters are shown in 
Fig.\ref{Fig8}. The parallel \hkl<1 0 0> dislocations and mixed parallel \hkl<1 
1 1>-\hkl<1 1 0>-\hkl<1 0 0> dislocations are the biggest. The \hkl <1 1 1> and 
the C15-like rings-\hkl<1 1 1> hybrid clusters also have large sizes. The 
latter is mostly due to the parallel component. The WSNAP potential shows large 
complex ring structures. \\

A density-based clustering algorithm to identify sub-cascades from primary 
damage states was developed \cite{BhardwajSubcascades}, which shows strong 
agreement with methods that use the peak state of the cascade to identify 
sub-cascades \cite{DEBACKER2021152887}. Fig.\ref{Fig9} quantifies sub-cascade 
counts across PKA energies for all potentials. A stiffness-dependent trend is 
observed at lower energies (<50 keV), wherein the tabGAP and DND-BN IAPs which 
are the stiffest, form significantly more sub-cascades whilst the JW, MS and 
WSNAP show reduced formation. Above 60 keV, tabGAP maintains elevated 
sub-cascade counts while DND-BN reverts to baseline behavior. The tabGAP 
potential shows a slight tendency toward increased sub-cascades for all PKA 
energies while the WSNAP potential consistently shows lesser sub-cascades 
count.

\section{Discussion \label{Discuss}}

The microstructural evolution of irradiated materials depends critically on 
both primary damage states and subsequent defect dynamics. This evolution is 
driven by defect migration, mutual interactions, and reactions with 
pre-existing microstructural features (e.g., grain boundaries). Our analysis 
reveals that defect counts across potentials remain statistically consistent 
(Fig. 2) regardless of stiffness (S) or stiffness-range ratios (S/R) when 
sampled across crystallographic directions. This supports established 
conclusions \cite{EffectZBLEDisp, TERENTYEV200665} that reasonable  stiffening 
parameters yield comparable defect counts. The above statement does not 
encompass the MS$_s$ potential from Sand et al \cite{MS} which has stiffness, 
S=-552 and range, R=-502 compared to the IAPs used in this study, which have S 
within the range -137 to -230 and a S/R values in the range -95 to -177. The 
MS$_s$ potential is much stiffer than the ZBL potential and the pair 
interaction potentials from two DFT simulations considering the 6s and 5d 
orbitals as valence electrons in one case and including the 5p electrons also 
as valence electrons in the other (Fig.1 from \cite{MS}). The MS$_s$ potential 
however illustrates the effect of an extremely stiff pair potential on a 
cascade. It is also known that the stiffness of an IAP affects cascade 
properties like volume of cascade and sphericity \cite{DEBACKER2021152887,MS}. 
For the IAP used in this study, the DND-BN and tabGAP potential have around 50 
$\%$ higher values for S (table.\ref{Table_2}) and this affects some of the 
results as discussed below. \\

\begin{figure}[h]
\centering{\includegraphics[width=\textwidth]{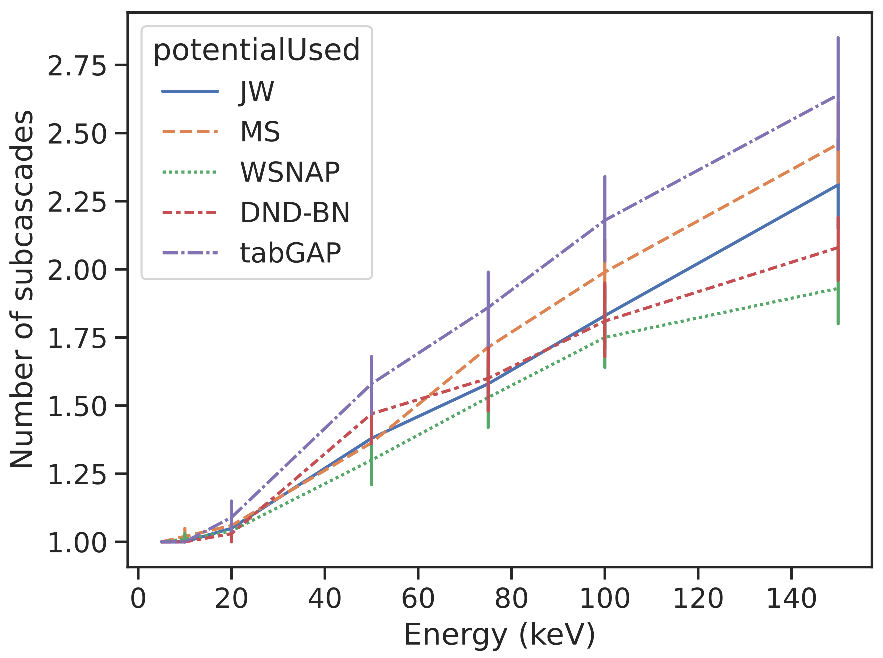}}
\caption{\label{Fig9} The number of sub-cascades as a function of the PKA
energy.}
\end{figure}

A pronounced inverse correlation between sub-cascade formation and total defect 
production is seen from fig.\ref{Fig2} and fig.\ref{Fig9} which reveals 
fundamental energy-partitioning mechanisms in cascade evolution. This manifests 
most dramatically in tabGAP, which has the highest summed SIA formation 
energies (rows 10-14 of Table.\ref{Table-3}), maximal sub-cascade counts 
(fig.\ref{Fig9}) and minumum defect yield (Fig. 2). From this it can be 
concluded that elevated defect formation energies suppress defect 
recombination, while extensive sub-cascade branching dissipates energy into 
lattice fragmentation rather than stable defect formation. Though tabGAP and 
DND-BN share high stiffness (S, Table.\ref{Table_2}), DND-BN's lower SIA 
formation energies enable greater defect survival. This explains their 
divergent defect counts despite similar repulsive core behavior. Similarly, a 
combination of least number of sub-cascades combined with a minimum value of 
vacancy formation energy amongst the 5 IAP seems to ensure that WSNAP has the 
largest number of vacancies in cluster in Fig.\ref{Fig5}. \\

While various interatomic potentials (IAPs) agree on the total number of 
defects created, they exhibit significant discrepancies in the fraction of 
defects forming clusters. As shown in Fig. \ref{Fig3}, the proportion of 
clustered defects varies by 20–40\% depending on the chosen IAP, consistent 
with prior observations \cite{DEBACKER2021152887}. These IAPs also reproduce 
similar fundamental classes of defect clusters, classified by orientation and 
morphology \cite{BhardwajClassify}; however, the distribution of clusters 
across these classes differs among IAPs, as evidenced in Fig. \ref{Fig7} and 
Fig. \ref{Fig8}. \\

Defect clustering depends on both the spatial distribution of displaced atoms 
at the cascade peak and their subsequent mutual interactions. The initial 
distribution is partly influenced by the stiffness of the IAP. Subsequent 
cluster formation (governed by defect formation energies) relies on the IAP’s 
behavior at intermediate interatomic distances, which is determined by the 
defect configuration data used during the IAP’s training. For instance, the 
formation energies of \hkl<1 1 0> and \hkl<1 1 1> defects critically influence 
the formation of C15-type clusters, as discussed previously. \\ 

Cluster stability depends not only on size but also on internal morphology 
\cite{Bhardwaj100Stability}. Sessile morphologies with high transition energies 
may act as nucleation sites for glissile defects during microstructural 
evolution, potentially leaving observable signatures in HRTEM experiments. 
Consequently, identifying defect cluster morphologies across different IAPs and 
comparing them with experimental data will aid in validating IAPs and improving 
predictive accuracy.

\section{Conclusions}
The primary damage characteristics from collision cascades in tungsten are 
investigated using five interatomic potentials (IAPs) for 100 randomly directed 
primary knock-on atoms (PKAs) at energies between 5 keV and 150 keV. Key 
findings include:

\begin{itemize}

\item All five IAPs produce defect clusters categorized into six distinct 
morphological/orientational types. While four IAPs predominantly generate 
\hkl<1 1 1> defect clusters, those exhibiting minimal differences in \hkl<1 1 
0> and \hkl<1 1 1> formation energies—coupled with low absolute \hkl<1 1 0> 
formation energies—promote C15-like and hybrid C15-like-\hkl<1 1 1> cluster 
formation. This tendency likely originates from liquid configurations included 
in these potentials' training datasets.

\item There is an inverse correlation between the number of defects formed 
and the number of sub-cascades. This, in conjunction with the defect formation 
energy for self interstitial atoms and various types of defects decides 
the total defect count.
 
\item Defect counts, cluster fractions, and sub-cascade numbers show no direct 
correlation with IAP stiffness (S) or stiffness/range (S/R) ratios. Instead, 
these ratios indirectly affect defect production through sub-cascade formation.
\end{itemize}

These results demonstrate that IAP selection critically impacts both defect 
clustering fractions and cluster morphologies. The \hkl<1 1 0>--\hkl<1 1 1> 
formation energy difference and the absolute value of \hkl<1 1 0> formation 
proves decisive for C15-ring-like cluster formation. While WSNAP exhibits 
significant deviation from experimental values for this parameter, DND-BN shows 
strong agreement. However, DND-BN displays limitations including negative 
thermal expansion and substantial deviations in elastic properties. We 
therefore recommend judicious IAP selection based on specific research 
objectives.

\section*{Acknowledgments}
We acknowledge the HPC team at CAD for $24 \times 7$ HPC maintenance and for
support in installing and running LAMMPS.

\section*{References}
\bibliographystyle{elsarticle-num}
\bibliography{MWarriers.bib}

\end{document}